\documentstyle[aasms4,flushrt]{article}

\lefthead{NAIK ET AL.}
\righthead{ABSOLUTE ASPECT OF BALLOON BORNE TELESCOPE}
\begin{document}

\title{Efficient absolute aspect determination
of a balloon borne far infrared telescope 
using a solid state optical photometer}

\author{M.V. Naik$^{1}$, S.L. D'Costa$^{2}$, S.K. Ghosh$^{3}$, B. Mookerjea$^{4}$, D.K. Ojha$^{5}$, R.P. Verma$^{6}$}
\affil{Tata Institute of Fundamental Research,  
Homi Bhabha Road, Colaba, 
Mumbai (Bombay) 400 005, India. \\
$^{1}$iraproj@lens.tifr.res.in, $^{2}$sdcosta@tifr.res.in, $^{3}$swarna@tifr.res.in, 
$^{4}$bhaswati@tifr.res.in, $^{5}$ojha@tifr.res.in, $^{6}$vermarp@tifr.res.in}

\authoraddr{ S. K. Ghosh \\
swarna@tifr.res.in \\
Department of Astronomy and Astrophysics \\
Tata Institute of Fundamental Research\\
Homi Bhabha Road, Colaba \\
Mumbai (Bombay) 400 005, India}

\begin{abstract}

 The observational and operational efficiency of
the TIFR 1 meter balloon borne far infrared telescope
has been improved by incorporating a multielement
solid state optical photometer (SSOP) at the Cassegrain 
focus of the telescope. The SSOP is based on a 
1-D linear photo diode array (PDA). The online and offline
processing schemes of the PDA signals which have been developed,
lead to improvement in the determination of absolute telescope 
aspect ($\sim$ 0\farcm8), which is very crucial for carrying out 
the observations as well as offline analysis. 
The SSOP and its performance during a recent balloon flight
are presented here.

\end{abstract}


\section{Introduction}

  The Tata Institute of Fundamental Research (TIFR)
1 meter balloon-borne far infrared (FIR) telescope is flown regularly 
to carry out observations of Galactic star forming regions, external
spiral galaxies etc (Daniel et al 1984; Bisht et al 1989). 
The orientation and pointing system
of this telescope uses a star tracker (ST) as a two axis 
angular position sensor (Almeida et al 1983; Ghosh \& Tandon 1982). 
A bright optical guide star, 
$m_{B} < 5$, within the field of view (2\arcdeg) of 
this ST, provides the positional reference.
Since the space angle between the nearest usable guide star 
and the FIR target ($\eta$), is typically $>$ 3\arcdeg\,
the star tracker is mechanically offset with respect to the 
main telescope about the two main control axes (viz., elevation \&
cross-elevation). The mechanical offset system allows for
$\pm$ 4\fdg5 motion about each axis (in steps of 
$\approx$ 20\arcsec). 
The mechanical offsetting of the ST is effected by  a pair
of stepper motor driven screws through trains of gears,
whose positions are measured by shaft encoders.
Although the pointing jitter of the telescope orientation system is 
$\approx$ 20\arcsec rms (adequate for observations at 200 $\mu$m, where
the diffraction limit is 50\arcsec), the achieved absolute positional accuracy
is only $\approx$ 2 -- 4\arcmin\ (for $\eta$ $>$ 3\arcdeg)
due to fabrication defects of mechanical components.
The above implies the necessity of in-situ absolute position calibration of the 
Cassegrain focal plane of the telescope.
In the past, a focal plane photomultiplier tube (FPPM) based optical photometer
(sensitive upto $m_{B}$ = 9.0, for 3 sec integration, typical of
observational rasters) has been used successfully
to improve the absolute aspect accuracy to $\sim 1\arcmin$
for $\eta$ as large as 5\arcdeg\ (Ghosh et al 1988, Das \& Ghosh 1991). 
With the introduction of
bolometer arrays in our two band (12 channel system; 
Verma, Rengarajan \& Ghosh 1993) 
FIR photometer, the use of the FPPM (which is effectively
a single pixel device) for achieving absolute
aspect, leads to very poor observational efficiency.
The multi element solid state optical
photometer (SSOP), is a solution which is briefly 
described in this paper. Sections 2 and 3 describe the 
SSOP and the relevant software processing schemes.
The results from a recent balloon flight, which quantify its performance,
are presented in Section 4.

\section{Solid State Optical Photometer}

 In order to achieve a good observational efficiency,
the FOV subtended by the entire detector-array of the FIR 
instrument must at least be covered by the SSOP.
The SSOP has to detect stars while FIR observations are in progress 
(e.g. the sky is chopped by wobbling the
secondary mirror at 10 Hz; and scanned at 0.5--1.0 arc min/sec).
These requirements translate to : resolution element $<$ 1\arcmin;
sensitivity of $ m_{R} \sim$ 10 (for
integration time corresponding to the typical raster scan);
and a dynamic range of  $\sim 10^{4}$.

The detector selected is an EG\&G Silicon Photodiode array, PDA-20-2, with
20 elements. Each element is 0.94 mm x 4.0 mm in size and the pitch is
1.0 mm resulting in a very small dead zone. It has a NEP 
of 7$ \times$ 10$^{-15}$ W Hz$^{\mathrm -1/2}$ (at +23\arcdeg\ C) 
and an operating temperature
range of +70\arcdeg\ C to -55\arcdeg\ C (ambient temperature at balloon
float altitude is $\approx$ -50\arcdeg\ C). 
Two consecutive elements are hardware ``binned"
(by connecting them in parallel) to implement an effective ``pixel" of
0\farcm87 (El) x 1\farcm7 (XEl) size, at the Cassegrain focal 
plane of the 1-meter (f/8) telescope.
Only 16 of the 20 elements of the PDA have been used (i.e. 8 pixels).
Hence, the used part of the PDA (1 pixel $\times$ 8 pixel array)
subtends an angle of 1\farcm7 $\times$ 6\farcm9 on the sky.
A two stage baffle 
with opening corresponding to $\approx$ f/7 precede the PDA. 
The PDA has reasonable spectral response from 5000 \AA ~ to 
10500 \AA ~ with a peak responsivity of 0.6 A/W at 9000 \AA .

The PDA is used in the photovoltaic mode. A bank of 
trans-impedance amplifiers, TIAs, pre-amplify
signals from each pixel (see Fig. 1), which are placed physically
close to the PDA inside a EMI insulated chamber. 
The preamplified signals are buffered and fed to
the 8 channel detector signal processing unit (DSPU).
Each DSPU channel consists of :
attenuator, buffer, composite band pass filters, phase sensitive 
detector (PSD), low pass filter and interface to the telemetry
system (see the DSPU block diagram in Fig. 2).
The final DSPU outputs from all 8 pixels are sampled at 10 Hz
and digitized (12 bit ADC) by the telemetry down-link.

 Since low frequency / DC drifts of the PDA signals 
are lost in PSD processing,
two selected pixels of the PDA are additionally processed 
through DC-coupled stages (with much lower gain to avoid
electronic saturation) and sampled at about 0.3 Hz. 
This is useful to monitor the background light level and
the dark current.

\section{Software for SSOP}

\subsection{Online processing}

 The PDA signals are processed online at the ground station,
while the telescope scans a pre-selected optical star in a clean field
near the far infrared target (within 20--30\arcmin\ ).
The results from this processing are used to
update the telescope model for absolute aspect.

 The signals from all the PDA pixels and the data from the
sensors relevant for the telescope aspect (all sampled at 10 Hz)
are stored in a time sequence for each scan line.
The time sequence of the PDA signals for each scan line are convolved with
a function which represents the PDA response for scan across a star (including
the effect of sky chopping). The time corresponding to the 
grand maximum of the convolved signal
sequence provides the telescope aspect corresponding
to the target star. The resulting aspects from several relevant scan lines
are combined to update / refine the existing model for the telescope
aspect.

\subsection{Off-line processing}
 
 The off line data processing involves
determination of the instantaneous telescope boresight using
the data from the two axis angular position (Star Tracker) and rate 
(Gyroscopes) sensors used in the telescope orientation and 
stabilization system (Ghosh et al 1988). 
The chopped SSOP signals (all 8 pixels) are gridded in
a two dimensional sky matrix (the two axes representing
the telescope coordinate system, viz., elevation \& cross elevation).
Signals from all 8 pixels of  SSOP are mixed using a 
focal plane model of their relative location, which is
determined during laboratory testings prior to the launch.
The telescope raster scans are parallel to the cross elevation axis.
The cell size used in this observed (chopped) signal matrix is 
0\farcm3 $\times$ 0\farcm3.
This observed signal matrix is
deconvolved using an indigeneously developed scheme using the
Maximum Entropy Method (MEM) similar to Gull
and Daniell, 1978 (see Ghosh et al., 1988, for details). 
The 2-D point spread function (PSF) used in the MEM scheme
is determined from the
scans across a bright star during the balloon flight.
The positions of the peaks in the deconvolved optical 
map represent detected stars which are
compared with various catalogues (SAO, HST Guide Star Catalogue etc)
to quantify any systematic shifts / effects.

\section{Performance of the SSOP}

  The SSOP system was flown during the balloon flight of the
1-meter far infrared telescope payload on March 8, 1998,
from Hyderabad, in central India.
The payload was at the float altitude of 31 km for 5.5
hours. During this flight, several bright stars
were scanned using SSOP (and sometimes using the FPPM)
to confirm the focal plane model and establish the absolute
aspect of the telescope. In addition, during the scans
across the FIR programme targets,
the SSOP has covered typically 600 square arc min of the
sky (Ghosh 1998). 

  The 2-D Point Spread Function (PSF) of the SSOP (corresponding
to one pixel) has been generated from the observations of the star $\rho$ Pup. 
The FWHM for a point source (BS 6546) after MEM deconvolution
is found to be 0\farcm85 $\times$ 1\farcm62 (Elev 
$\times$ Cross-elev), which is very close to the expected
value.

  The off-line processing has been carried out for 9
mapped regions, each covering about 30\arcmin $\times$
25\arcmin\ area. Figure 3 shows the resulting optical isophot contour
map from a typical observation.
The brightest and the faintest star in this map correspond to 
$m_{R}$ of 7.06 and 9.76 respectively.
Clear detections of well identified stars are marked on this map. 
A total of 40 stars have been detected and identified in
these 9 fields. The final
absolute map coordinates are determined from the 
shift parameters ($\Delta$RA, $\Delta$Dec) which best align 
the peaks of the map with the coordinates of identified stars. 
For the present sample of 9 regions mapped, the shift angle
(${\theta}_{corr} = \sqrt{\Delta RA^{2} + \Delta Dec^{2}}$)
is found to be increasing with, $\eta $, the
offset angle between the telescope axis and the Star Tracker axis.
The RA and Dec components of the residual angles ($res_{\alpha}$,
$res_{\delta}$) show a gaussian distribution (see Fig. 4). 
The standard deviations 
($\sigma (res_{\alpha})$ = 44\arcsec; $\sigma (res_{\delta})$ = 
29\arcsec) reflect the ultimate absolute aspect errors in the final maps 
and quantify the cumulative effects of pointing jitter;
electronic and data processing noise; and the
quality of telescope optics (the primary \& secondary mirrors
are designed for FIR wavelengths and hence are
very poor for optical wavelengths). 
By dividing the entire sample into two categories on the basis
of the offset angle $\eta $ ($\eta > 2^{0}$; \& the rest), 
it has been found that 
$\sigma(res_{\alpha})$ \& $\sigma(res_{\delta})$
are not sensitive to $\eta$.
Hence, using the SSOP, it has been possible to achieve an 
absolute aspect accuracy of $\approx$ 0\farcm8, in the
presence of mechanical imperfections leading to 1.5--4\arcmin\
errors.

  The spectral response of the PDA elements is such that
the $m_{R}$ magnitude of stars represents the
signal expected from the SSOP. 
A total of 22 stars, for which $m_{R}$ could be found / estimated
from the literature, have been used to calibrate the SSOP and quantify the
system linearity and sensitivity. The SSOP is found to be linear
within the testable range of $2 < m_R < 9.7$.
The faintest star detected corresponds to $m_{R}$ = 10.9 in our sample.
The expected sensitivity for the SSOP (for identical observational 
conditions) is $m_R$ = 10.0. Hence, the achieved sensitivity is quite
close to its design goal.

 The analysis of the DC coupled channels (2 of the 8 pixels)
implies a large increase in the scattered light background near the telescope
focal plane during the time when moon (illuminated fraction
= 0.82) was above horizon.
The maximum observed background corresponds to 
$\approx$ 14.9 mag ($m_{R}$) arcsec$^{-2}$.

\section{Conclusions}

 A multielement solid state optical photometer (SSOP) has been
developed and successfully used at the Cassegrain 
focal plane of the TIFR 1-meter balloon borne far infrared
telescope. This SSOP has been used on-line as well as
off-line to achieve higher absolute positional accuracy (0\farcm8)
of the telescope during a balloon flight.
The achieved sensitivity of the SSOP corresponds to 
the stellar magnitude of $m_{R}$ $\sim$ 10.0
(for typical raster scans used for FIR targets)
which is consistent with the expectations.
The SSOP has also improved the observational and operational
efficiency of the telescope.

\vskip 1cm
\centerline{\it Acknowledgements}
\vskip 0.5cm
It is a pleasure to thank members of the Infrared
Astronomy Group of TIFR for their encouragement and support.



%

%
\newpage

\figcaption[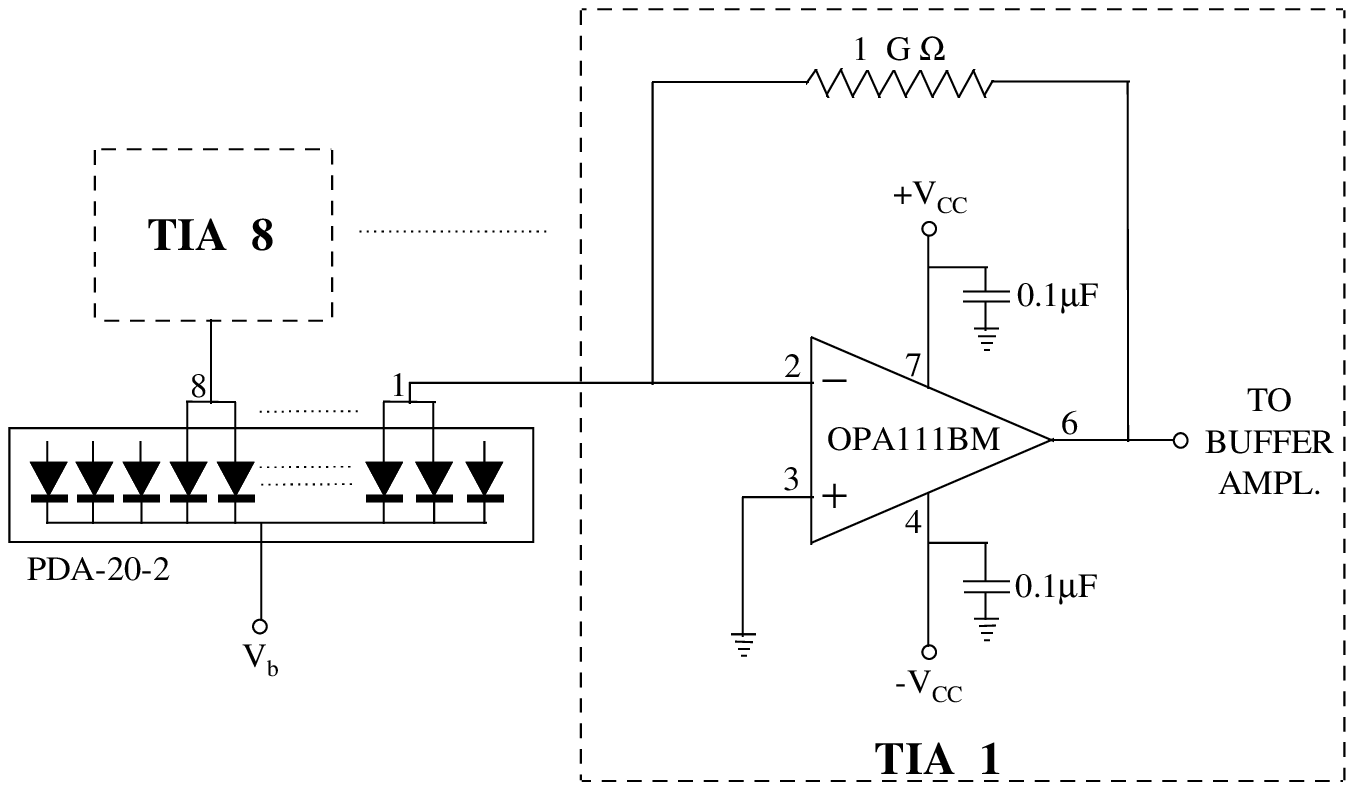] 
{The Trans-Impedance Amplifier (TIA) used to
preamplify the signals of the Photo Diode Array (PDA) elements.
Each pixel comprises of two PDA elements, and 8 pixels 
have been used for the Solid State Optical Photometer (SSOP).
\label{}
}

\figcaption[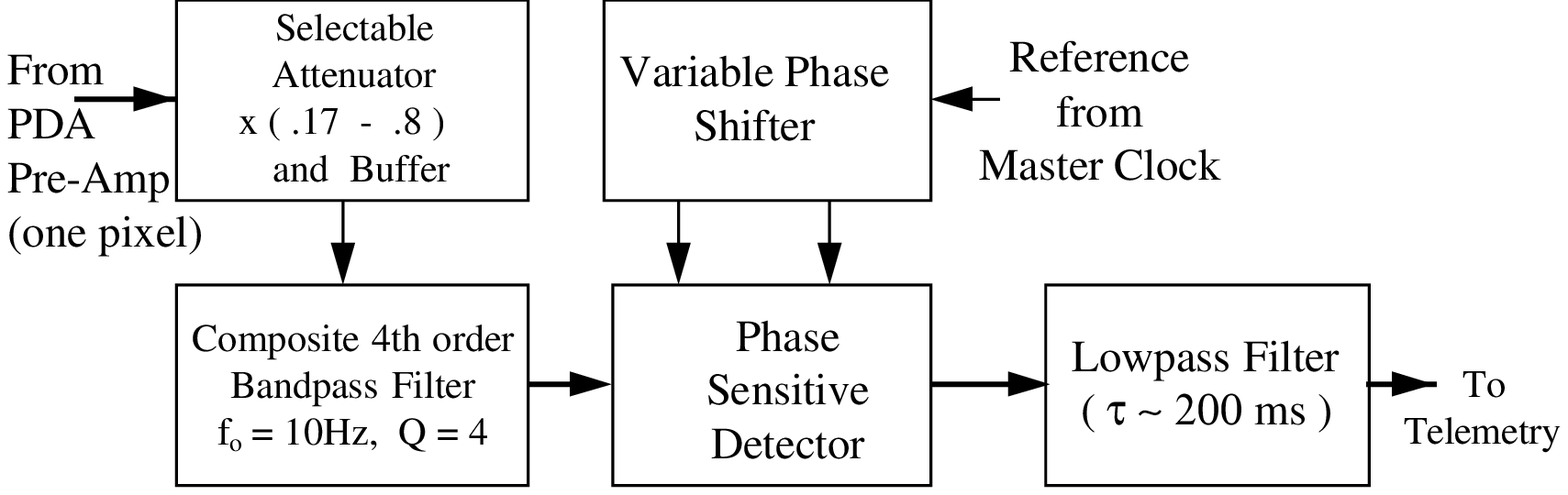]
{Block diagram of the on-board Detector Signal Processing 
Unit (DSPU) of the SSOP.
\label{}
}

\figcaption[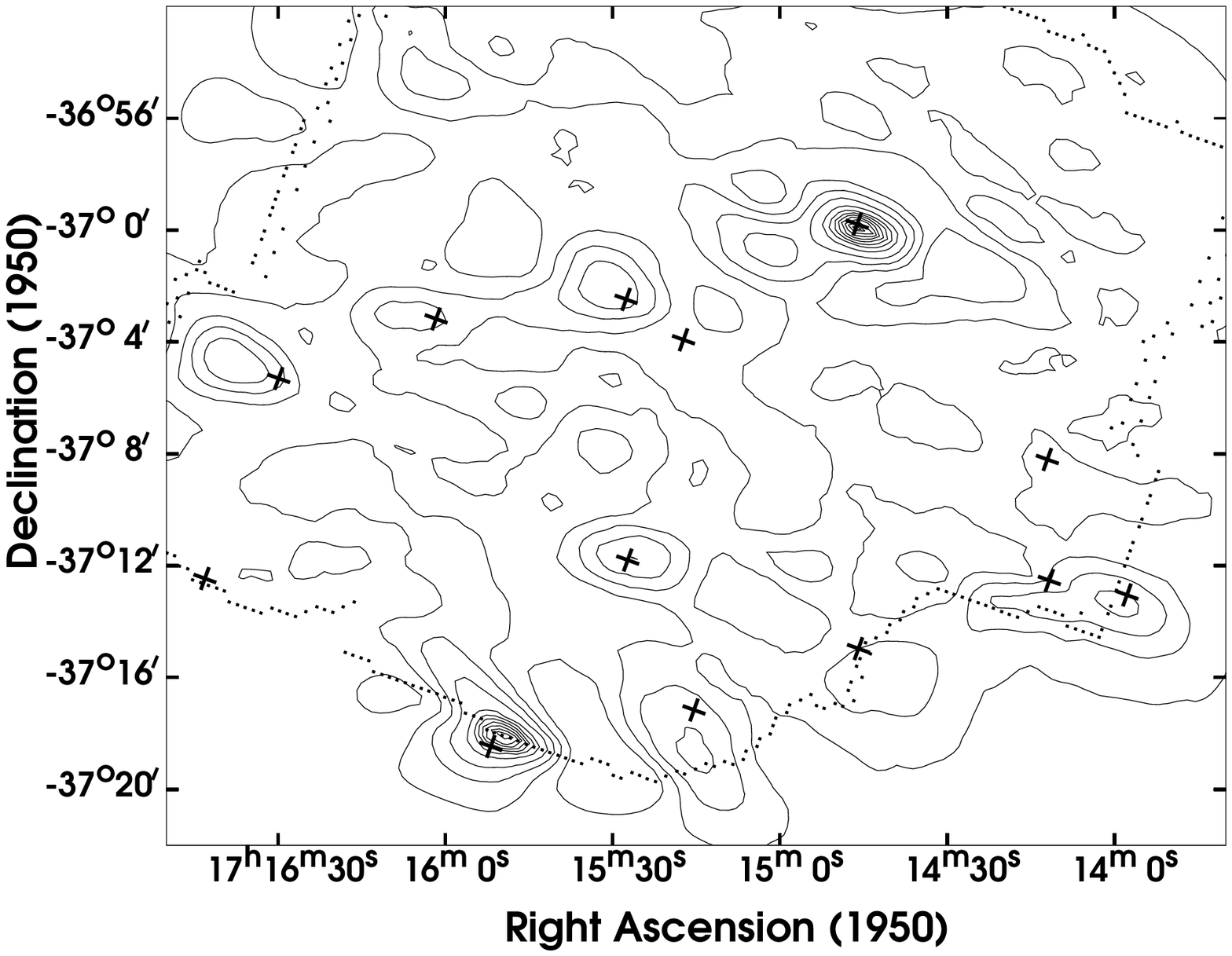]
{
In-flight SSOP optical map of the region near IRAS 17160-3707.
The contour levels are 95\%, 90\%, 80\%, 70\%, 60\%, 50\%,
40\%, 30\%, 20\%, 10\%, 5\%, 2.5\% and 1\% of the peak intensity, which is
= 7.67 mag ($m_{R}$) arcmin$^{-2}$.
The symbol $+$ represents detected and identified stars; and the 
dots mark the boundary of the region mapped. 
\label{}
}

\figcaption[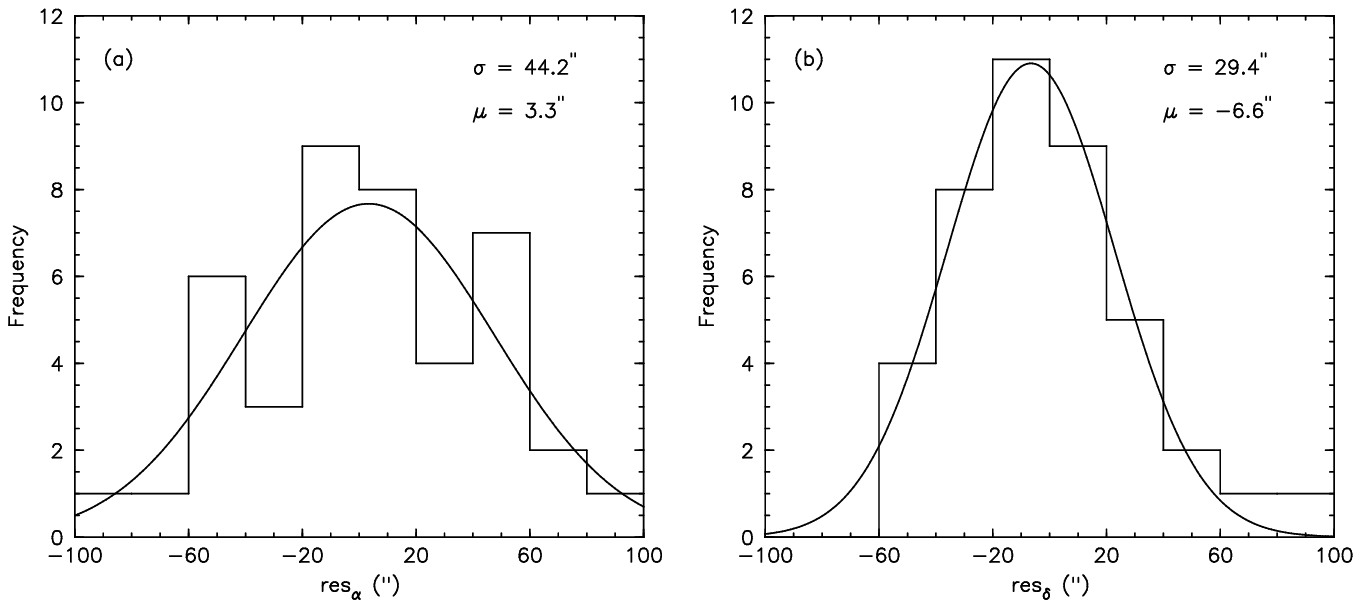]
{The distributions of the (a) RA and (b) Dec components of 
the residuals ($res_{\alpha}$ \& $res_{\delta}$), after 
applying the correction (see text). The smooth curves denote the
best fit gaussians. The means ($\mu$) are approximately zero as expected
and the standard deviations ($\sigma$) quantify the achieved absolute 
aspect accuracy.
\label{}
}


\end{document}